\documentclass[
  aps,  
  twocolumn,
  prl,
  superscriptaddress,
  floatfix
]{revtex4}

\usepackage{amssymb,amsfonts,amsmath,dsfont}
\usepackage{times} 
\usepackage{graphicx} 
\usepackage{gensymb}
\usepackage{pifont}
\usepackage{bm} 
\usepackage{color}
\usepackage{accents}
\usepackage{mathtools}
\usepackage{hyperref}
\usepackage{url}
\hypersetup{colorlinks=true, citecolor=blue, urlcolor=blue, linkcolor=blue}
\usepackage{multirow}
 
\makeatletter
\newcommand*{\supplementarystart}{%
  \close@column@grid%
  \clearpage%
  \onecolumngrid%
  \setcounter{enumiv}{0} % resets counter for references
  \setcounter{equation}{0} % resets counter for equations
  \setcounter{figure}{0} % resets counter for figs
  \setcounter{table}{0} % resets counter for tables
  \setcounter{page}{1}
  \c@secnumdepth=4
  \renewcommand{\theequation}{s\arabic{equation}} % equations numbered with S...
  \renewcommand{\bibnumfmt}[1]{[s##1]} % bibtems [S...]
  \renewcommand{\@onlinecite}{s\citealp} % citations [S...]
  \renewcommand{\cite}[1]{{[}\onlinecite{##1}{]}}
  \renewcommand{\thefigure}{s\arabic{figure}}
  \renewcommand{\thetable}{s\Roman{table}}
  \renewcommand{\thepage}{s\arabic{page}}
}
\makeatother
 
\newcommand{\s}{\sum\limits} 
 
\newcommand{\pa}{\partial} 
 
\newcommand{\be}{\begin{equation}} 
\newcommand{\e}{\end{equation}} 
\newcommand{\beml}{\begin{subequations}} 
\newcommand{\eml}{\end{subequations}} 
\newcommand{\beq}{\begin{eqnarray}} 
\newcommand{\eq}{\end{eqnarray}} 
\newcommand{\ba}{\begin{array}} 
\newcommand{\ea}{\end{array}} 
\newcommand{\bpm}{\begin{pmatrix}} 
\newcommand{\epm}{\end{pmatrix}} 
\newcommand{\bc}{\begin{cases}} 
\newcommand{\ec}{\end{cases}} 
\newcommand{\lt}{\left} 
\newcommand{\rt}{\right} 
\newcommand{\n}{\nonumber} 
\newcommand{\la}{\langle} 
\newcommand{\ra}{\rangle}

\newcommand{\bb}{\mathbf} 
\newcommand{\h}{^\dagger}

\DeclareMathOperator{\diag}{diag}

\begin{document}
	
\title{Non-collinear ground state from a four-spin chiral exchange in a tetrahedral magnet}	
\author{I.\,A.~Ado}
\affiliation{Institute for Molecules and Materials, Radboud University, Heyendaalseweg 135, 6525AJ Nijmegen, The Netherlands}	
\author{O.~Tchernyshyov}
\affiliation{Institute for Quantum Matter and Department of Physics and Astronomy, Johns Hopkins University, Baltimore, MD 21218, USA}	
\author{M.~Titov}
\affiliation{Institute for Molecules and Materials, Radboud University, Heyendaalseweg 135, 6525AJ Nijmegen, The Netherlands}	

\date{\today}

\begin{abstract}
We propose a quartic chiral term $m_x m_y m_z \nabla \cdot \mathbf{m}$ for the energy density of a cubic ferromagnet with broken parity symmetry (point group $T_d$). We demonstrate that this interaction causes a phase transition from a collinear ferromagnetic state to a non-collinear magnetic cone ground state provided its strength exceeds the geometric mean of magnetic exchange and cubic anisotropy. The corresponding non-collinear ground state may also be additionally stabilized by an external magnetic field pointing along certain crystallographic directions. The four-spin chiral exchange does also manifest itself in peculiar magnon spectra and favors spin waves with the wave vector that is perpendicular to the average magnetization direction.  
\end{abstract}
	
\maketitle
Conduction electrons are capable of mediating magnetic interactions of localized spins in a magnet. The resulting indirect symmetric magnetic exchange, known as Ruderman--Kittel--Kasuya--Yosida (RKKY) interaction \cite{RudermanKittel,Kasuya,Yosida}, stands, for example, behind the giant magnetoresistance effect \cite{Grunberg1989,Fert1988}. In metallic or semiconducting magnets with broken inversion symmetry and strong spin-orbit interaction of conduction electrons, the same RKKY mechanism is responsible for an indirect long-range asymmetric exchange \cite{Smith1976,FertLevy1980}. On the level of Ginzburg-Landau micromagnetic energy, the asymmetric exchange from such a ``twisted'' RKKY interaction becomes indistinguishable from the Dzyaloshinskii-Moriya interaction (DMI) \cite{Moriya1960,Dzyaloshinsky1958} and is represented by terms that are linear with respect to the spacial gradients of magnetization vector: the so-called Lifshitz invariant (LI) terms \cite{LI1941}. Nowadays, electron-mediated indirect asymmetric exchange between a pair of localized spins is commonly referred to as the DMI. 

The DMI is responsible for non-collinear long-range magnetic order, such as the helical spin-density waves \cite{Ishikawa1976}. Indeed, the presence of linear-in-gradient terms in micromagnetic energy may often make a collinear (ferromagnetic or anti-ferromagnetic) order unstable with respect to formation of cone, helix, vertex or skyrmion crystals. Such structures are indeed observed in many conducting magnets or magnetic multilayers with broken inversion symmetry such as MnSi, FeGe, Ir/Co/Pt, or Pt/CoFeB/MgO \cite{Ishikawa1976,Shirane1983,Muhlbauer2009,Grigoriev2009,Moreau-Luchaire2016,Woo2016} to name a few. 

The presence of spin-orbit interaction is necessary but not sufficient condition for the appearance of a finite electron-mediated DMI. For example, for electrons yielding a two dimensional Rashba model with quadratic electron dispersion, the effect of conduction electrons on localized spins is finite only in half-metal regime, while it becomes identically zero irrespective of the Rashba spin-orbit interaction strength if both spin-split sub-bands are occupied \cite{AdoDMI2018}. 

Nevertheless, it has been recently shown that spin-orbit interaction may not only lead to a finite DMI, but may also produce finite and even larger contributions to the free energy from electron-mediated multi-spin exchange \cite{AdoDMI2020}. Such multi-spin interactions do contribute in the linear order with respect to magnetization gradients. We refer to these contributions as chiral multi-spin interactions. 

The symmetry analysis performed by Bogdanov and Yablonskii \cite{BogdanovYablonskii1989} establishes a correspondence between the point group symmetry of a crystal and the combinations of Lifshitz invariants that may arise in the micromagnetic energy functional to describe DMI. 

More recently, it has been shown by Ado \textit{et al.} \cite{AdoDMI2020} that there exist three specific point groups for crystals with broken inversion symmetry: $T_d$, $D_{3h}$ and $C_{3h}$ for which all LI terms in micromagnetic functional are forbidden, while multi-spin chiral exchange interactions are allowed by symmetry.  

In this Letter we propose a four-spin chiral exchange interaction in a tetrahedral magnet -- the one characterized by the point group $T_d$. We demonstrate that such interaction leads to a peculiar magnon dispersion and to instability with respect to the formation of the helical ground state.
 
Our analysis might apply to B20 compounds, half-metal halcogenides, pyrochlores and Heusler alloys including Cu$_3$FeTe$_4$, Lu$_2$V$_2$O$_7$, Cr$_\textrm{x}$Zn$_\textrm{1-x}$Te, Mn$_\textrm{x}$Zn$_\textrm{1-x}$S and many related materials. The multi-spin chiral interaction is expected to be strong in conducting magnets with large spin-orbit interaction of charge carriers.   

%%%%%%%%%%%%%%%%%%%%%%%%%%%%%
\begingroup
\begin{table}
\begin{center}
{\renewcommand{\arraystretch}{1.5}
\tabcolsep=5pt
\begin{tabular}{c|c|c|}
& $2$-spin (LI) &  $4$-spin (non-LI) \\
\hline
$O$ & $\bb{m}\cdot(\nabla \times \bb{m})$ & 
\begin{tabular}{cc}   
\multirow{2}{*}
& $\sum_\alpha m^3_\alpha (\nabla \times \bb{m})_\alpha$ \\
& $m_x^2\mathcal L^{(x)}_{y x} + m_y^2\mathcal L^{(y)}_{z x}+m_z^2\mathcal L^{(z)}_{x y}$ \\
\end{tabular}
\\
\hline
$T_d$ & None & $m_x m_y m_z (\nabla \cdot {\bb m})$ \\
\hline
\end{tabular}
}
\end{center}
\caption{\label{table} Energy density from two-spin (LI) and four-spin (non-LI) chiral interactions in the point group $T$ (chiral tetrahedral symmetry) that is a common subgroup of $O$ (chiral octahedral symmetry) and $T_d$ (full tetrahedral symmetry). The notation $\mathcal{L}^{(\gamma)}_{\alpha\beta} = m_\alpha \partial_\gamma m_\beta - m_\beta\partial_\gamma m_\alpha$ denotes the Lifshitz invariant (LI).
} 
\label{tab:Ttable}
\end{table}
\endgroup
%%%%%%%%%%%%%%%%%%%%%%%%%%%%%

%%%%%%%%%%%%%%%%%%%%%%%%%%%%
%%%% fig:state
%%%%%%%%%%%%%%%%%%%%%%%%%%%%
\begin{figure*}[t]
\centerline{\includegraphics[width=1.8\columnwidth]{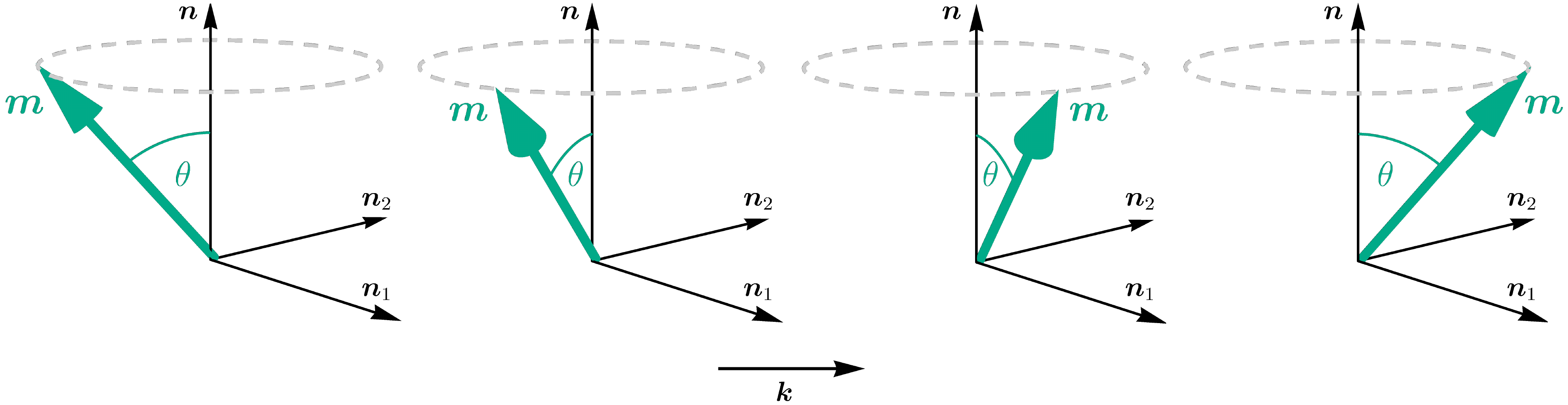}} 
\caption{Schematic illustration of the magnetic cone state that minimize the energy of Eq.~(\ref{energy0}). The state wave-vector is perpendicular to the average magnetization, $\bb{k}\cdot \bb{n}=0$, that is characteristic for the 4-spin chiral interaction $w_\textrm{4S}\propto m_xm_ym_z\bb{\nabla}\cdot\bb{m}$.}
\label{fig:state}
\end{figure*}
%%%%%%%%%%%%%%%%%%%%%%%%%%%%%

We describe a ferromagnet with the micromagnetic energy functional $E[\bb{m}]$ that depends on a dimensionless magnetization vector field $\bb{m}(\bb{r})$ of the unit length. Indirect chiral exchange interactions are represented in this functional by terms that are linear in magnetization gradients. Such terms may only arise if the underlying magnetic lattice lacks the inversion symmetry. 

In the Table~\ref{tab:Ttable} we list the results of the symmetry analysis for a lattice with the point group $T$ (chiral tetrahedral symmetry). This is a common subgroup of the point groups $O$ and $T_d$. One can readily see that the two-spin chiral interaction, the bulk DMI with the energy density $w_\textrm{DMI} \propto \bb{m}\cdot(\bb{\nabla}\times\bb{m})$, arises in the point group $O$ but not in the point group $T_d$. 

As any $2$-spin chiral interaction, the bulk DMI is represented by a particular combination of Lifshitz invariants:  $w_\textrm{DMI} \propto \mathcal{L}^{(x)}_{yz}+\mathcal{L}^{(y)}_{zx}+\mathcal{L}^{(z)}_{xy}$, where $\mathcal{L}^{(\gamma)}_{\alpha\beta} = m_\alpha \partial_\gamma m_\beta - m_\beta\partial_\gamma m_\alpha$.  The key role of the bulk DMI $w_\textrm{DMI}$ on the formation of helical spin density waves has been known since the early theory works \cite{Baryahtar1969,Nakanishi1980,Bak1980,LANDAUbook}. This interaction is responsible for skyrmion crystal and helical spin phases in MnSi, MnFeSi, FeCoSi, FeGe and in many other magnetic materials \cite{Ishikawa1976,Shirane1983,Muhlbauer2009,Grigoriev2009}. At the same time, if spin-orbit induced splitting of conduction electron bands becomes comparable with the $s$-$d$ exchange energy one may also expect 4-spin (and in general multi-spin) chiral interactions to play an important role \cite{AdoDMI2020}. 

From the Table~\ref{tab:Ttable} one can also see that there exist two possible $4$-spin chiral interactions in the point group $O$ and only one in the point group $T_d$, where all 2-spin chiral terms are forbidden. 

It is worth noting that the four-spin chiral interaction term, $w_\textrm{4S}\propto m_xm_ym_z\bb{\nabla}\cdot\bb{m}$ was missed in the classification presented in Ref.~\cite{AdoDMI2020} since the latter has been restricted to multi-spin terms in the form of products of Lifshitz invariants and magnetization-dependent coefficients \cite{Karin}. The term $w_\textrm{4S}$  cannot, however, be written in such a form. 

Thus, we can formulate a universal energy functional of $T_d$ ferromagnet, $E[\bb{m}]= \int d^3\bb{r}\, \lt[w(\bb{r}) -\bb{H}\cdot\bb{m}\rt]$, where $\bb{H}$ stands for external magnetic field measured in energy units, while the energy density of the magnet reads, 
\be
\label{Model}
w= A\s_\alpha (\bb{\nabla}m_\alpha)^2+8B\,m_xm_ym_z\bb{\nabla}\cdot\bb{m}+K\s_{\alpha}m_\alpha^4,
\e
where we collected all possible terms up to the forth order in magnetization.  

The first term in Eq.~(\ref{Model}) represents the usual symmetric exchange, $A>0$, the second term corresponds to the new $4$-spin chiral interaction discussed above, and the last term is the cubic anisotropy. Note that we include the first anisotropy constant ($K_1=-2K$) but ignore terms of the $6$-th order in magnetization ($K_2=0$) \cite{Landau8}. Throughout the Letter we assume that the ferromagnet is kept well below Curie temperature, hence $|\bb{m}|=1$. 

Even though the 4-spin chiral interaction has never been previously considered, it is not difficult to see that it leads to instability of collinear order towards formation of magnetic cone, provided the anisotropy is sufficiently weak.

%%%%%%%%%%%%%%%%%%%%%%%%%%%%
%%%% fig:PhD
%%%%%%%%%%%%%%%%%%%%%%%%%%%%
\begin{figure*}[t]
\centerline{\includegraphics[width=1.7\columnwidth]{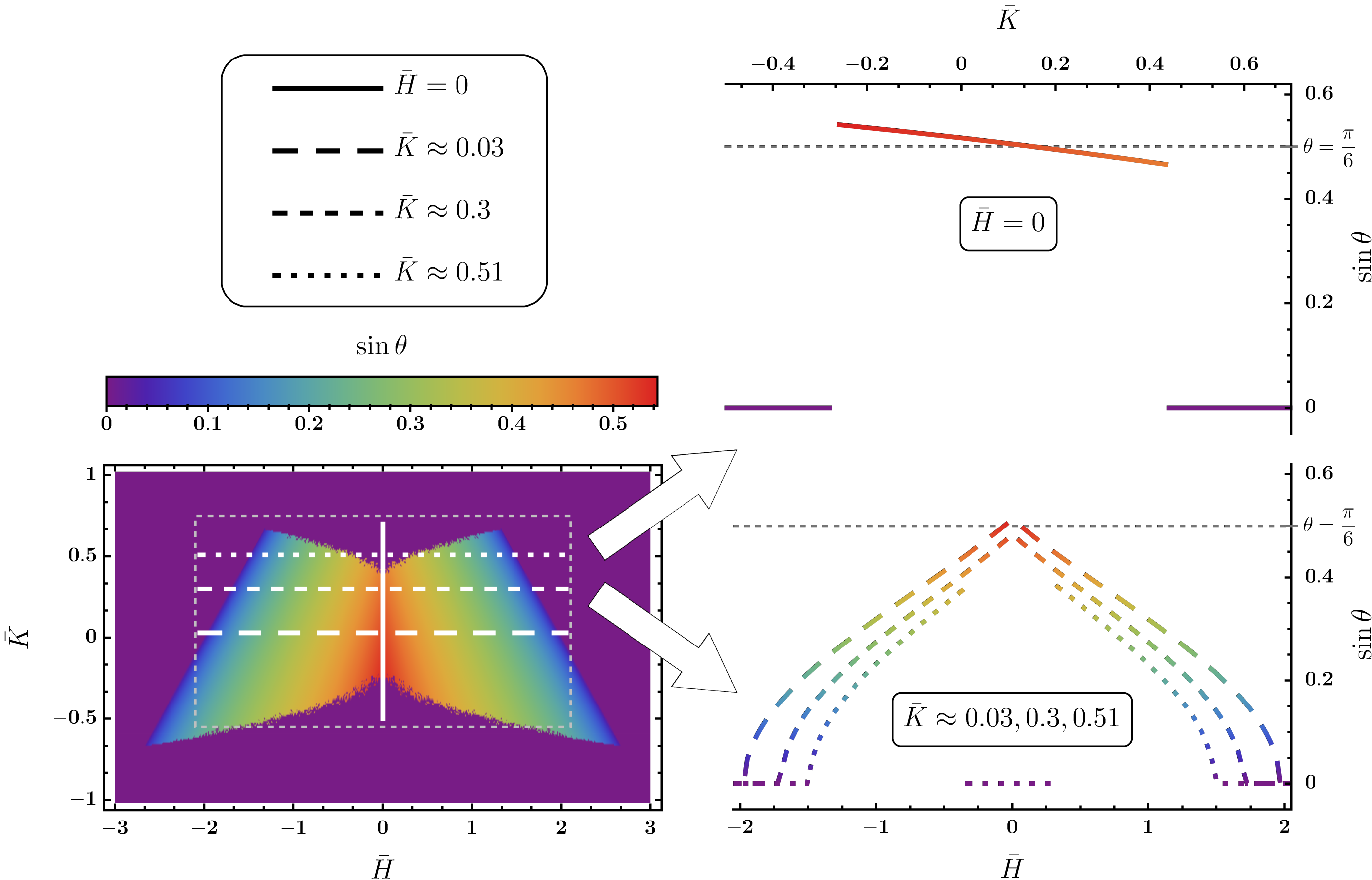}} 
\caption{The color plot is obtained by numerical minimization of the function $\mathcal{E}(\bb{k},\bb{n},\theta)$ of Eq.~(\ref{energy0}) and represents the value of $\sin\theta$ (the span of magnetic cone) at the global minimum,  provided external magnetic field is directed as $\bar{\bb{H}}=\bar{H}(0,1,1)/\sqrt{2}$. Non-collinear magnetic cone state (finite $\theta$ and $\bb{k}$) is realized for moderate values of $\bar{K}$ and $\bar{H}$. The upper left panel shows the horizontal crosssection with $\bar{H}=0$, while the lower left panel shows three vertical crosssections for $\bar{K}=0.03$, $0.3$, and $0.51$. The angle $\theta$ smoothly deviates from zero across the lines $\bar{K}=2-|\bar{H}|$, which correspond to the second order phase transition. Noisy borders for $\bar{K} \approx \pm 0.5$ correspond to the first order phase transition from collinear to a non-collinear state with a finite $\theta$. The corresponding jumps are also seen in the left panels.}
\label{fig:PhD}
\end{figure*}
%%%%%%%%%%%%%%%%%%%%%%%%%%%%%

In order to see the instability of the collinear state, let us consider a generalized spiral ansatz for magnetization vector, 
\be
\label{ansatz}
\bb{m}(\bb{r})= \bb{n}\cos\theta + \lt[\bb{n}_1\cos{(\bb{k}\cdot\bb{r})} + \bb{n}_2\sin{(\bb{k}\cdot\bb{r})}\rt]\sin\theta 	,
\e
where $\bb{n}_1$, $\bb{n}_2$ and $\bb{n}=\bb{n}_1\times \bb{n}_2$ are mutually orthogonal unit vectors; the wave vector reversal, $\bb{k}\to -\bb{k}$, is equivalent to $\bb{n}_2\to -\bb{n}_2$ (helicity reversal); $\theta=0$ corresponds to a collinear state, while $\theta=\pi/2$ corresponds to a pure helix.

A translation $\bb{r} \mapsto \bb{r} + \Delta \bb{r}$ is equivalent to a rotation of the reference frame through the angle $\bb{k} \cdot \Delta \bb{r}$ about the $\bb{n}$ direction. Translational symmetry therefore implies the existence of a Goldstone mode involving the rotation of spins about $\bb{n}$.
 
We further substitute Eq.~(\ref{ansatz}) into Eq.~(\ref{Model}) and average the result over the phase $\bb{k} \cdot \Delta \bb{r}$ to obtain a Landau energy density $\mathcal{E}=E/V$. The latter becomes a function of the parameters $\bb{k}$, $\bb{n}$ and $\theta$ of the conical state (\ref{ansatz}) 
\begin{align}
\mathcal{E}=&A\,k^2\sin^2\!\theta - B\,\bb{k}\cdot\bb{v}(\bb{n})\sin^2\!\theta\,(1-5\cos^2\!\theta)\n\\
&+K\lt[u_1(\theta)+u_2(\theta)c(\bb{n})\rt]- \bb{n}\cdot\bb{H}\,\cos\theta,
\label{energy}
\end{align}
where we introduced 
\beml
\begin{align}
\label{vdef}
\bb{v}(\bb{n})= &(n_x(n_y^2-n_z^2), n_y(n_z^2-n_x^2), n_z(n_x^2-n_y^2)),\\
\label{cdef}
c(\bb{n})=&3(n_y^2n_z^2+n_z^2n_x^2+n_x^2n_y^2),\\
u_1(\theta)=&\cos^4\!\theta+(3/4)\sin^4\!\theta,\\
u_2(\theta)=&2\cos^2\!\theta\sin^2\!\theta-(2/3)\cos^4\!\theta-(1/4)\sin^4\!\theta.
\end{align}
\eml
Note that the transverse polarization condition $\bb{n}\cdot\bb{v}(\bb{n})=0$ follows directly from Eq.~(\ref{vdef}).

The four-spin interaction sets the energy scale $B^2/A$ that defines the non-collinear order. After rescaling 
\be
\bb{k}=B\bar{\bb{k}}/A,\quad \bb{H}=B^2\bar{\bb{H}}/A,\quad K=B^2\bar{K}/A, 
\e
one can rewrite the energy density of Eq.~(\ref{energy}) as follows:
\be
\mathcal{E} =\frac{B^2}{A}\lt[(\bar{\bb{k}}-\bar{\bb{k}}_0)^2\sin^2\!\theta+u({\bb{n},\theta})-\bb{n}\cdot\bar{\bb{H}}\cos\theta \rt],
\label{energy0}
\e
where $\bar{\bb{k}}_0=\bb{v}(\bb{n})\; (1-5\cos^2\theta)/2$ is a characteristic wave-vector and $u({\bb{n},\theta})= \bar{K}\lt[u_1(\theta)+u_2(\theta)c(\bb{n})\rt]-\bar{k}^2_0\sin^2\!\theta$ is an effective potential. 

The energy density (\ref{energy0}) has an absolute minimum either in a collinear state with $\bb{k}=0$ or in a conical state with $\bb{k}=\bb{k}_0= B\bar{\bb{k}}_0/A$.  The wavevector $\bb{k}_0$ is always perpendicular to $\bb{n}$ as follows from the condition $\bb{v}\cdot\bb{n}=0$. For the non-collinear phase, the resulting conical magnetic order is illustrated schematically in Fig.~\ref{fig:state}. This is in contrast to the bulk DMI $\propto \bb{m}\cdot(\bb{\nabla}\times\bb{m})$ that stabilizes conical or helical states with $\bb{k}_0$ parallel to $\bb{n}$. One can see that the span of magnetic cone $\theta$ may, at best, only slightly exceed the value $\pi/6$, while the pure helix, $\theta=\pi/2$, is never reached.  

The energy density (\ref{energy0}) is obtained within the ansatz of Eq.~(\ref{ansatz}) and may not represent the absolute minimum of the micromagnetic energy. It is, however, known that the same ansatz of Eq.~(\ref{ansatz}) is often very accurate, e.\,g. for the case of bulk DMI (the point group O). We may, therefore, hope that the minimization of energy in Eq.~(\ref{energy0}) does reflect the true minimization of the original micromagnetic energy functional (\ref{Model}). 

The result of numerical energy minimization in Eq.~(\ref{energy0}) is illustrated in Fig.~\ref{fig:PhD} by plotting the dependence of $\sin\theta$ on both $\bar{K}$ and $\bar{H}$ at the absolute energy minimum. 

For zero field and small anisotropy, $-0.28 < \bar{K} < 0.44$, we find a non-collinear conical state with $\bb{k}=\bb{k}_0$ and $\theta \approx \pi/6$. The minimum is reached for $\bb{n}=(0,1,1)/\sqrt{2}$, $\bb{v}=(0,1,-1)/2\sqrt{2}$, and for the other 11 equivalent directions of $\bb{n}$ that are related by the rotation symmetries of the $T_d$ point group (see Table~\ref{tab:GS} of the Supplemental material \cite{supplementary}). 

In the limit of large anisotropy, the ground state is collinear. For example, for zero field one finds the minimal energy density, $\mathcal{E}=K/3$ for $\bar{K}> 0.44$ with the magnetization along a body diagonal such as $\bb{n}=(1,1,1)/\sqrt{3}$, and $\mathcal{E}=K$ for $\bar{K}<-0.28$ with the magnetization along $\bb{n}=(0,0,1)$ and symmetry equivalents. An external magnetic field applied in $\la 011\ra$ (or any equivalent) direction can additionally stabilize the non-collinear state as can be indeed seen in Fig.~\ref{fig:PhD}. 

Generally, the angle $\theta$ deviates smoothly from zero across the lines $\bar{K}=2-|\bar{H}|$ indicating a second order phase transition. The noisy borders of the color plot in Fig.~\ref{fig:PhD} correspond to the first order transition that is characterized by the competition of minima at finite $\theta$ and $\theta=0$ (see also the left panels). 

Let us now investigate how the four-spin chiral interaction may affect the magnon spectra. To that end we linearize Landau-Lifshitz equation $\pa \bb{m}/\pa t= \bb{H}_\textrm{eff}\times \bb{m}$ with respect to a small variation $\delta\bb{m}$. We consider a collinear phase, where the unit vector $\bb{n}$ yields the equation $(\bb{H}-4K \bb{n}^{o3})\times\bb{n}=0$ with $\bb{n}^{o3}=(n_x^3, n_y^3,n_z^3)$.  Instead of solving the resulting cubic equation we introduce the Lagrange multiplier $\lambda=\lambda(\bb{H},K)$ that is set by the algebraic equation
\be
\label{effective}
\bb{H}_\textrm{eff}= \bb{H}-4K \bb{n}^{o3}-\lambda\bb{n}=0,
\e
alongside with two independent components of the vector $\bb{n}$.

Using the ansatz $\bb{m}=\bb{n}+\delta\bb{m}\,\exp(i\omega_{\bb{q}} t-i\bb{q}\cdot\bb{r})$ with $\bb{n}\cdot\delta\bb{m}=0$, we, then, obtain the magnon dispersion \cite{supplementary}
\be
\label{general}
\omega_{\bb{q}} = \sqrt{(\Omega_q+ 4 c K)^2 +16 K^2(d^2 -c^2)} - 8 B\, \bb{v}\cdot\bb{q},
\e
where $\Omega_q=2A\,q^2+\lambda$, $\bb{v}$ and $c$ are defined in Eqs.~(\ref{vdef}) and (\ref{cdef}), correspondingly, and $d=3\sqrt{3}\, n_x n_y n_z$. 

For $H\gg |K|$, one finds $\bb{n}=\bb{H}/H$, hence $\lambda=H$ and Eq.~(\ref{general}) is reduced to  
\be
\label{largeH}
\lt.\omega_{\bb{q}}\rt|_{H\gg K} = 2A\,(\bb{q}-\bb{q}_0)^2+H-8B^2v^2/A,
\e
where $\bb{q}_0=(2B/A)\bb{v}$. 

The vector $\bb{q}_0$, which defines the effect of the four-spin chiral exchange, takes on a particular direction that is orthogonal to $\bb{n}$ by construction. This is again in sharp contrast to the effect of the bulk DMI for which $\bb{q}_0\propto \bb{n}$.

It is worth noting that the coefficient $B$ does not enter the magnon dispersion in the absence of external field. Indeed, for $\bb{H}=0$, the ground state magnetization $\bb{n}$ is set by the sign of the anisotropy constant only. 

For $K>0$ one finds $\bb{n}=(1,1,1)/\sqrt{3}$, which corresponds to $\lambda=-4 K/3$, $\bb{v}=0$, $c=d=1$. Therefore, the magnon dispersion reads $\omega_q=2A\,q^2+ 8K/3$. 

For $K<0$ one finds $\bb{n}=(0,0,1)$, $\lambda=-4K$, $\bb{v}=0$, $c=d=0$, hence $\omega_q=2A\,q^2+ 4|K|$. 

To maximize the effect of the four-spin term one needs to drive the length of the vector $\bb{v}$ to its maximal value $v=1/2$. This can be achieved again by applying an external field in a direction $\la 011\ra$ or in any other equivalent crystallographic direction. 

The coefficient $B$ can be quantified by measuring the difference $\delta\omega_{\bb{q}}=\omega_{\bb{q}}-\omega_{-\bb{q}}=-16 B\bb{v}\cdot\bb{q}$ for the wave-vector $\bb{q}$ that is orthogonal to the magnetization direction $\bb{n}$, provided the vector $\bb{v}$ is finite. 

Thus, the new interaction term in cubic crystals with broken inversion symmetry does lead to non-reciprocal magnon dispersion. Similarly to the bulk DMI, it breaks the symmetry with respect to the wave vector reversal $\bb{q}\to -\bb{q}$, but in a direction of $\bb{q}$ that is orthogonal to magnetization. The bulk DMI leads to $\bb{q}\to -\bb{q}$ non-reciprocity in the direction parallel to magnetization. 

It is evident from Eqs.~(\ref{general},\ref{largeH}) that the four spin chiral interaction shifts the minimum of magnon energy $\bb{q}\propto \bb{q}_0$. Moreover, the results suggest that the frequency $\omega_{\bb{q}}$ becomes negative at least for $H\simeq 2B^2/A$, provided anisotropy is sufficiently weak, $|K|\lesssim B^2/A$. Such negative values of $\omega_{\bb{q}}$ are unphysical and indicate an instability of the collinear order. Low-energy magnons in the presence of non-collinear periodic ground state form a banded spectrum that we do not analyze in this Letter. 

So far we have discussed the 4-spin chiral interaction in the continuum theory limit. One possible Heisenberg equivalent of this interaction can be constructed on a pyrochlore lattice. Let us consider the four vertices of a regular tetrahedron with coordinates $\mathbf r_0 = (0,0,0)$, $\mathbf r_1 = (0,-a/4,-a/4)$, $\mathbf r_2 = (-a/4,0,-a/4)$, and $\mathbf r_3 = (-a/4,-a/4,0)$, where $a$ is the cubic lattice constant of the pyrochlore lattice. We further define the four unit vectors pointing from the center of the tetrahedron to the respective sites: 
\beq
\bb{n}_0 = (+1,+1,+1)/\sqrt{3}, && \bb{n}_1 = (+1,-1,-1)/\sqrt{3}, \n \\
\bb{n}_2 = (-1,+1,-1)/\sqrt{3}, && \bb{n}_3 = (-1,-1,+1)/\sqrt{3},\quad
\eq
which satisfy $\bb{n}_i \cdot \mathbf n_j =(4 \delta_{ij}-1)/3$. 

With these notations, the four-spin chiral exchange interaction is given by the following energy: 
\beq
U_4  &=& (\mathbf n_0 \cdot \mathbf S_0)
    (\mathbf e_x \cdot \mathbf S_1)
    (\mathbf e_y \cdot \mathbf S_2)
    (\mathbf e_z \cdot \mathbf S_3)
\n \\
&+& (\mathbf e_x \cdot \mathbf S_0)
    (\mathbf n_1 \cdot \mathbf S_1)
    (-\mathbf e_z \cdot \mathbf S_2)
    (-\mathbf e_y \cdot \mathbf S_3)
\n \\
&+& (\mathbf e_y \cdot \mathbf S_0)
    (-\mathbf e_z \cdot \mathbf S_1)
    (\mathbf n_2 \cdot \mathbf S_2)
    (-\mathbf e_x \cdot \mathbf S_3)
\n \\
&+& (\mathbf e_z \cdot \mathbf S_0)
    (-\mathbf e_y \cdot \mathbf S_1)
    (-\mathbf e_x \cdot \mathbf S_2)
    (\mathbf n_3 \cdot \mathbf S_3),
\label{eq:U4}
\eq
where $\mathbf e_\alpha$ stand for the unit vectors in the chosen coordinate frame, $\alpha=x,y,z$, while $\mathbf S_i$ stand for spins on respective lattice cites. The gradient expansion of $U_4$ to the lowest order, 
\be
\mathbf S_i(\mathbf r_i) = S\lt[
\mathbf m(0) 
+ \left. 
    (\mathbf r_i \cdot \nabla) \mathbf m(\mathbf r) 
\right|_{\mathbf r = 0}
+ \ldots \rt],
\e
and subsequent integration by parts yields the chiral 4-spin term with $B=-a/8S^4$. 

In conclusion, we suggest the existence of the four-spin indirect magnetic interaction that may be responsible for the appearance of a long-range non-collinear magnetic order in ferromagnets with magnetic lattice yielding $T_d$ point group symmetry.  We demonstrate that the usual DMI interaction on such a lattice does not contribute to the micromagnetic energy functional (in the linear order with respect to magnetization gradients) and cannot cause an instability of the collinear order, while the four-spin chiral interaction can. A similar situation may arise in crystals with $D_{3h}$ and $C_{3h}$ point group symmetries that are rather common among two dimensional magnets. Thus, taking into account possible four-spin chiral exchange interactions is important for understanding non-collinear magnetic order in these systems. 

{\it Acknowledgments} --- The authors are thankful to Arne Brataas, Karin Everschor-Sitter, Helena Gomonay and Marcos Guimaraes for illuminating discussions. The authors acknowledge support from the JTC-FLAGERA Project GRANSPORT. M.T. and O.T. performed part of this work at the Kavli Institute for Theoretical Physics supported by the US NSF Grant PHY-1748958. O.T. has been supported by the US DOE Basic Energy Sciences, Materials Sciences and Engineering Award DE-SC0019331.
	
\bibliographystyle{apsrev4-1}
\bibliography{Biblio}

\begin{thebibliography}{26}
\bibitem [{Kar()}]{Karin}%
  \BibitemOpen
  \href@noop {} {}\bibinfo {note} {We are indebted to Karin Everschor-Sitter
  for pointing this out to us.}\BibitemShut {Stop}%
\end{thebibliography}

%================Supplementary===============
\supplementarystart

\centerline{\bfseries\large ONLINE SUPPLEMENTAL MATERIAL}
\vspace{6pt}
\centerline{\bfseries\large Non-collinear ground state from a four-spin chiral exchange in a tetrahedral magnet}
\vspace{6pt}
\centerline{I.\,A.~Ado, O.\,A.~Tchernyshyov, and M.~Titov}
\begin{quote}
In this Supplemental Material we provide some additional technical details that may further clarify the main text of the Letter. 
\end{quote}

\maketitle

\section{Energy minimization}

For $K=0$ and $\bb{H}=0$, the energy density of Eq.~(\ref{energy0}) of the main text reads 
\be
\label{Fenergy}
\mathcal{E}= \frac{B^2}{A}\lt[(\bar{\bb{k}}-\bar{\bb{k}}_0)^2\sin^2\theta + u({\bb{n},\theta})\rt], 
\e
where the effective potential is 
\be
u({\bb{n},\theta})=-\frac{1}{4}v^2 \sin^2\theta (1-5\cos^2\theta)^2.
\e
Therefore, the minimum of $\mathcal{E}$ is reached for $\bb{k}=\bb{k}_0$ (non-collinear state), for the direction of $\bb{n}$ that maximize the length of the vector $\bb{v}$, and for the angle $\theta =\arcsin(\sqrt{4/15})\approx 0.543 \approx \pi/6$ that maximize the function $\sin^2\theta (1-5\cos^2\theta)^2$. 

The quantity $v^2$ is maximal for 12 equivalent crystalographic directions $\bb{n}$ of the average magnetization. These specific directions are listed in the Table \ref{tab:GS} together with the corresponding vector $\bb{v}=(n_x(n_y^2-n_z^2), n_y(n_z^2-n_x^2),n_z(n_x^2-n_y^2))$. 

%%%%%%%%%%%%%%%%%%%%%%%%%%%%%
\begingroup
\begin{table}[b]
\begin{center}
{\renewcommand{\arraystretch}{1.5}
\tabcolsep=5pt
\begin{tabular}{c c | c c}
$\sqrt{2}\bb{n}$ &  $2\sqrt{2}\bb{v}$ & $\sqrt{2}\bb{n}$ &  $2\sqrt{2}\bb{v}$ \\
\hline
$(0,1,1)$ & $(0,1,-1)$ & $(0,1,-1)$ & $(0,1,1)$ \\
$(1,0,1)$ & $(-1,0,1)$ & $(-1,0,1)$ & $(1,0,1)$ \\
$(1,1,0)$ & $(1,-1,0)$ & $(1,-1,0)$ & $(1,1,0)$ \\
$(0,-1,1)$ & $(0,-1,-1)$ & $(0,-1,-1)$ & $(0,-1,1)$\\
$(1,0,-1)$ & $(-1,0,-1)$ & $(-1,0,-1)$ & $(1,0,-1)$\\
$(-1,1,0)$ & $(-1,-1,0)$ & $(-1,-1,0)$ & $(-1,1,0)$\\
\end{tabular}
}
\end{center}
\caption{Possible choices of the vectors $\bb{n}$ and $\bb{v}$ that define 12 equivalent non-collinear states (related by the $T_d$ point group symmetry rotations) for $\bb{H}=0$. Such states minimize the energy in Eq.~(\ref{energy}) of the main text for sufficiently small anisotropy parameter as shown in Fig.~\ref{fig:PhD} of the main text. Note that $\bb{n}\cdot \bb{v}=0$ for each pair of vectors. 
} 
\label{tab:GS}
\end{table}
\endgroup
%%%%%%%%%%%%%%%%%%%%%%%%%%%%%

For a finite $K$ there exists a competition between the non-collinear minimum at $\bb{k}=\bb{k}_0$ and the collinear minimum at $\bb{k}=0$. The first one always corresponds to 12 directions of $\bb{n}$ specified in the Table \ref{tab:GS}, where the particular value of the angle $\theta$ now depends on $K$. The collinear minimum corresponds to $\theta=0$ and two possible sets of $\bb{n}$ as discussed in the main text. For $K>0$ it is given by $\bb{n}=(1,1,1)/\sqrt{3}$ and equivalent directions, while for $K<0$ it is given by $\bb{n}=(0,0,1)$ and equivalent directions. 

Applying external field in one of the 12 directions $\bb{n}$ listed in the Table \ref{tab:GS} may naturally increase the range of the values of $K$ that correspond to a non-collinear ground state as illustrated in Fig.~\ref{fig:PhD} of the main text.

The transition between collinear and non-collinear ground state is the competition between two local minima of $F$ that are separated by a potential barrier. Hence, it is always the first order phase transition. 

\section{Magnon dispersion}

New interaction term in cubic crystals with broken inversion symmetry does also affect the spin-wave (or magnon) dispersion in a non-trivial way. Similarly to the bulk DMI, the four-spin chiral interaction does break the symmetry with respect to magnon wave vector reversal $\bb{q}\to -\bb{q}$, but in the direction of $\bb{q}$ that is orthogonal to magnetization. The bulk DMI leads to  $\bb{q}\to -\bb{q}$ symmetry breaking in the direction parallel to magnetization. 

To find the magnon dispersion we consider the functional $E[\bb{m}]= \int d^3\bb{r}\, \lt[w(\bb{r})-\bb{H}\cdot\bb{m}\rt]$, where $\bb{H}$ is an external field. The corresponding Landau-Livshitz (LL) equation has the form
\be
\frac{\pa\bb{m}}{\pa t} = -\bb{m}\times\bb{H}_\textrm{eff},
\e 
where the effective field is given by $\bb{H}_\textrm{eff}=-\delta E[\bb{m}]/\delta\bb{m}$. We restrict ourselves to the case of collinear ground state. In this case one can linearize the LL equation using the simple ansatz
\be
\bb{m}(\bb{r},t)= \bb{n}+\delta\bb{m}\,e^{i\omega_{\bb{q}}t-i\bb{q}\cdot\bb{r}},
\e
where the vector $\bb{n}$ minimizes the functional $E[\bb{m}]$. It is important that $|\bb{n}|=1$ and $\bb{n}\cdot\delta\bb{m}=0$. 

Consequently, the vector $\bb{n}$ yields the algebraic equation
\be
\label{min1}
(\bb{H} - 4K \bb{n}^{o3})\times\bb{n}=0, 
\e
where $\bb{n}^{o3}=(n_x^3, n_y^3,n_z^3)$. 

Instead of analyzing the solutions of Eq.~(\ref{min1}), we introduce the Lagrange multiplier $\lambda$ and rewrite Eq.~(\ref{min1}) in the form of Eq.~(\ref{effective}) of the main text, $\bb{H}_\textrm{eff}= \bb{H}-4K \bb{n}^{o3}-\lambda\bb{n}=0$, that has to be solved for two independent components of the unit vector $\bb{n}$ and the parameter $\lambda$. Among many solutions one has to pick up those that minimize the energy. 

The dispersion relation for magnons can be expressed via the vector $\bb{n}$. To find this dispersion it is convenient to parameterize $\delta\bb{m}=\bb{n}\times \bb{\delta}$, where $\bb{\delta}$ is some vector. Obviously, the condition $\bb{n}\cdot\delta\bb{m}=0$ is, then, automatically fulfilled. 

The linearized LL equation takes a relatively simple matrix form 
\be
\lt[i\omega_{\bb{q}} \hat{\mathcal{M}} - \hat{\mathcal{A}}\rt]\bb{\delta}=0, \qquad \mbox{with}\quad 
\hat{\mathcal{A}}=(2 A\, q^2+\lambda)\, \hat{\mathcal{M}}^2 - 8 i B\, \bb{v}\cdot\bb{q}\, \hat{\mathcal{M}}  + 12 K\, \hat{\mathcal{M}}\hat{\mathcal{M}}_0^2\hat{\mathcal{M}},
\e
where we introduced the matrices
\be
\hat{\mathcal{M}} = \bpm 0 & - n_z & n_y \\ n_z & 0 & -n_x & \\ - n_y & n_x & 0 \epm,\qquad \mbox{and}\quad 
\hat{\mathcal{M}}_0= \bpm n_x &  0 & 0 \\ 0 & n_y & 0 \\ 0 & 0 & n_z \epm,
\e
and the notations
\beml
\begin{align}
\bb{v}&= (n_x(n_y^2-n_z^2),n_y(n_z^2-n_x^2),n_z(n_x^2-n_y^2)),\\
c&=3 (n_x^2n_y^2+n_y^2n_z^2+n_z^2n_x^2),\\
d&=3\sqrt{3}\, n_xn_yn_z.
\end{align}
\eml

The projection of the vector $\bb{\delta}$ on the direction of $\bb{n}$ is irrelevant and corresponds to zero eigenvalue of the matrix $i\omega_{\bb{q}} \hat{\mathcal{M}} - \hat{\mathcal{A}}$ (which is obvious since $\hat{\mathcal{M}}\bb{n}=0$). Two non-trivial eigenvalues of the matrix $(i\omega_{\bb{q}} \hat{\mathcal{M}} -\hat{\mathcal{A}})$ read 
\be
\label{Delta}
\Delta_\pm(\omega_{\bb{q}})= 2Aq^2+\lambda+4cK \pm  \sqrt{(\omega_{\bb{q}}+8 B \bb{v}\cdot\bb{q})^2+ 16(c^2-d^2)K^2}.
\e
The corresponding eigenvectors $\bb{\delta}_\pm$ belong to the plane that is orthogonal to the vector $\bb{n}$. 

The equation on magnon dispersion takes the form
\be
\Delta_+(\omega_{\bb{q}})\, \Delta_-(\omega_{\bb{q}})=0,
\e
which is solved by $\omega_{\bb{q}}=\omega^{\pm}_{\bb{q}}$ with
\be
\omega^{\pm}_{\bb{q}} = \pm \sqrt{(2Aq^2+\lambda+4 c K)^2 +16K^2 (d^2-c^2)} -8 B\bb{v}\cdot\bb{q}, 
\e
where $\omega^-_{\bb{q}}=-\omega^+_{-\bb{q}}$. The solution $\omega^+_{\bb{q}}$ corresponds to Eq.~(\ref{general}) of the main text. 

It is also convenient to use the dimensionless variables $\bar{\bb{q}}$, $\bar{\bb{H}}$ and $\bar{K}$, defined by the substitutions
\be
\bb{q}=\frac{B}{A}\bar{\bb{q}},\qquad K=\frac{B^2}{A}\bar{K}, \qquad \lambda=\frac{B^2}{A}\bar{\lambda}, .
\e
In terms of these variables one writes
\be
\label{dimensionless}
\omega^+_{\bb{q}}=\frac{4B^2}{A}\lt[\sqrt{\lt(\bar{q}^2/2+\bar{\lambda}/4 + c \bar{K}\rt)^2+\bar{K}^2(d^2-c^2)}-2\bb{v}\cdot\bar{\bb{q}}\rt].
\e

\begin{figure}[t]
\centerline{\includegraphics[width=0.4\columnwidth]{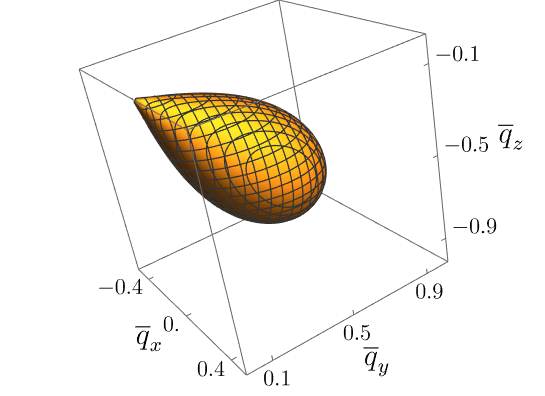}}
\caption{The gapless surface, $\omega_{\bb{q}}=0$, computed from Eq.~(\ref{general}) with $K = B^2/A = H/2$ and $\bb{H}=H 
 (0,1,1)/\sqrt{2}$ for dimensionless momenta $\bar{\bb{q}}=A\bb{q}/B$.}
\label{fig:gapless}
\end{figure}

In Fig.~\ref{fig:gapless} we use Eq.~(\ref{dimensionless}) to visualize a zero-frequency surface, $\omega^+_{\bb{q}}=0$, in $\bb{q}$ space for the choice $K = B^2/A$, $H=2 B^2/A$ and $\bb{H}=H(0,1,1)/\sqrt{2}$. The magnon dispersion of Eq.~(\ref{dimensionless}) (and, equivalently, of Eq.~(\ref{general}) of the main text) still holds for the wave-vectors $\bb{q}$ laying far outside the volume enclosed by the surface. Corresponding magnons have wave lengths that are much shorter than the period of a non-collinear state and are, therefore, not sensitive to the instability. 

The dispersion of Eq.~(\ref{dimensionless}) becomes, however, unphysical for the wave-vectors $\bb{q}$ approaching the surface. To find the magnon dispersion in this case it is necessary to take into account the non-collinear character of the ground state. 

%%%%%%%%%%%%%%%%%%%%%%%%%%%%
%%%% fig:gapless
%%%%%%%%%%%%%%%%%%%%%%%%%%%%
%%%%%%%%%%%%%%%%%%%%%%%%%%%%%

\section{Magnon polarization}

It is also worthwhile to compute two non-trivial eigenvectors of the matrix $i\omega_{\bb{q}} \hat{\mathcal{M}} - \hat{\mathcal{A}}$. To do that it is convenient to rotate the matrix $\hat{\mathcal{M}}$ into a basis where it is diagonal,
\be
V\h\hat{\mathcal{M}}V=\diag(0,i,-i).
\e
After such a rotation one finds
\be
\label{repr}
\hat{\mathcal{D}}=V\h\lt(i\omega_{\bb{q}} \hat{\mathcal{M}} - \hat{\mathcal{A}}\rt)V=
\bpm 0 & 0 & 0\\   0 & X-Y & Z e^{i\phi} \\  0 & Z e^{-i\phi} & X+Y \epm, 
\e
where 
\be
X=2Aq^2+\lambda+4cK,\qquad Y=\omega_{\bb{q}} +8 B\, \bb{v}\cdot\bb{q}, \qquad Z=4K\sqrt{c^2-d^2},
\e
while the angle $\phi$ is non-universal and depends on a particular choice of $V$.  In particular, one can always choose the rotation matrix $V$ such that $\phi=0$. 

The eigenvalues of the matrix $\hat{\mathcal{D}}$ are given by $\Delta_\pm=X\pm\sqrt{Y^2+Z^2}$  that obviously coincide with those of Eq.~(\ref{Delta}). Magnon spectra are defined by $\Delta_-(\omega_{\bb{q}})=0$, hence $Y^2=X^2-Z^2$. 

In Eq.~(\ref{repr}) one can explicitly see the non-trivial $2\times 2$ sub-space that corresponds to the plane perpendicular to $\bb{n}$. The eigenvalues $\Delta_\pm$ of the matrix $V\h\lt(i\omega_{\bb{q}} \hat{\mathcal{M}} - \hat{\mathcal{A}}\rt)V$ are evidently given by Eq.~(\ref{Delta}).

The corresponding eigenvectors of the matrix $V\h\lt(i\omega_{\bb{q}} \hat{\mathcal{M}} - \hat{\mathcal{A}}\rt)V$ can be written, up to a factor, as
\be
\tilde{\bb{\delta}}_+\propto \bpm 0\\ 2Aq^2+\lambda+4cK-(\omega_{\bb{q}}+8 B\, \bb{v}\cdot\bb{q})-\Delta_- \\ 4K\sqrt{c^2-d^2} e^{-i\phi} \epm, 
\qquad
\tilde{\bb{\delta}}_-\propto \bpm 0\\ 4K\sqrt{c^2-d^2} e^{i\phi}\\ \omega_{\bb{q}}+8 B\, \bb{v}\cdot\bb{q}-(2Aq^2+\lambda+4cK)+\Delta_- \epm.
\e

%When discussing magnon spectra we are interested in frequencies that yield $\Delta_-(\omega_{\bb{q}})=0$. 

%Thus, one finds two eigenvectors $\bb{\delta}_\pm=V \tilde{\bb{\delta}}_\pm$ of the matrix 

\end{document}